\documentclass{nature}
\usepackage{amsmath}
\usepackage{lineno}%line number

\usepackage{graphicx}
\title{Interlayer fractional quantum Hall effect in a coupled graphene double-layer}

\author{Xiaomeng Liu$^{1}$, Zeyu Hao$^1$, Kenji Watanabe$^2$, Takashi Taniguchi$^2$, Bertrand Halperin$^1$ \& Philip Kim$^1$}

\begin{document}

\maketitle

\begin{affiliations}
 \item Department of Physics, Harvard University, Cambridge, Massachusetts 02138, USA
 \item National Institute for Material Science, 1-1 Namiki, Tsukuba 305-0044, Japan
\end{affiliations}

\begin{abstract}

In two-dimensional (2D) electron systems under strong magnetic fields, interactions can cause fractional quantum Hall (FQH) effects \cite{Tsui1982, Laughlin1983}. Bringing two 2D conductors to proximity, a new set of correlated states can emerge due to interactions between electrons in the same and opposite layers \cite{Halperin1983,Chakraborty1987,Eisenstein2014,Jain2007}. Here we report interlayer correlated FQH states in a system of two parallel graphene layers separated by a thin insulator. Current flow in one layer generates different quantized Hall signals in the two layers. This result is interpreted by composite fermion (CF) theory\cite{Jain1989} with different intralayer and interlayer Chern-Simons gauge-field coupling. We observe FQH states corresponding to integer values of CF Landau level (LL) filling in both layers, as well as ``semi-quantized" states, where a full CF LL couples to a continuously varying partially filled CF LL. Remarkably, we also recognize a quantized state between two coupled half-filled CF LLs, attributable to an interlayer CF exciton condensate.

\end{abstract}

The energy levels of a non-interacting 2D electron system in a magnetic field are quantized into a discrete set of LLs with degeneracy proportional to the area of the system\cite{Jain2007}. A key parameter in these systems is the LL filling factor $\nu = n \phi_0 / B$, where $n$ is the electron density, $B$ is the magnetic field perpendicular to the layer, and the magnetic flux quantum $\phi_0 = h/e $, with $-e$ the electron charge. Integer quantized Hall (IQH) effects occur when $\nu$ is an integer, where the Fermi level is in an energy gap between two LLs, and Coulomb interactions between electrons can often be ignored. However, Coulomb interactions have a dominant effect in partially filled LLs, lifting the LL degeneracy and causing new collective states of matter to appear at a certain set of fractional values of $\nu$, which is known as the fractional quantum Hall (FQH) effect\cite{Tsui1982}.

In single-layer systems, the most commonly observed FQH states can be understood in terms of the composite fermion (CF) picture\cite{Jain1989}. Here, the electrons are each bound to even number ($2m$) of quanta of an emergent Chern-Simons (CS) gauge field to form CFs, leaving only relatively weak residual interactions between them. Since the CS field combines with the applied magnetic field, the CFs experience an effective magnetic field $B^* = B - 2 m n \phi_0$, which is generally weaker than the original field $B$.  From this effective magnetic field, we can relate $\nu$ to the CF LL filling factor $p$: $\nu = p/(2mp +1)$.   
If $p$ is an integer, positive or negative, then the CF system is predicted to have an energy gap, and the electrons will be in a corresponding FQH state, with $\nu$ equal to a fraction with odd denominator. Because of this energy gap, the FQH state has vanishing longitudinal electric resistance $R_{xx}$ and quantized Hall resistance\cite{Tsui1982} $R_{xy}=h/\nu e^2$. FQH states also have quasiparticles with fractional charge and fractional quantum statistics (anyons), different from the statistics of bosons or fermions\cite{Laughlin1983,Halperin1984}.

The scope of quantum Hall physics further expands when we bring two layers close to each other, allowing strong Coulomb coupling between them, while suppressing a direct interlayer tunneling. One much-studied state in such systems is the interlayer-coherent IQH state, which was first observed for the total filling factor $\nu_{tot} = 1$, where 
$\nu_{tot} \equiv\nu_{top}+\nu_{bot}$ is the sum of the filling factors in the top and bottom layers\cite{Eisenstein2014,Suen1992a,Eisenstein1992a}. Due to the Coulomb interaction, electrons in one layer are correlated with holes in the other. The ground state may be described as having a Bose condensate of interlayer excitons, added to a starting state in which one layer is empty while the other has a completely full Landau level.  As the density of excitons can be varied continuously, the interlayer coherent state can exist over a wide range of values for the difference in layer occupations, while $\nu_{tot}$ is fixed to be an integer. 

Several experimental methods, including Coulomb drag\cite{Kellogg2002,Liu2017,Li2017}, counterflow\cite{Kellogg2004,Tutuc2004a}, and tunneling measurements\cite{Spielman2000}, have been exploited to demonstrate the interlayer correlation, superfluidity and coherence of exciton condensation. In Coulomb drag measurements, current $I$ is driven through one of the layers (drive layer), while the other layer (drag layer) is electrically not connected. When tunneling is absent, development of a large drag voltage $V_{drag}$ proportional to $I$ provides a direct evidence for strong interlayer correlation. For $\nu_{tot}=1$, the Hall drag resistance $R_{drag}^{xy}=V_{drag}^{xy}/I$ is quantized to the same value as Hall resistance of the drive layer, $R_{drive}^{xy}=R_{drag}^{xy}=h/e^2$, proving interlayer correlation and exciton superfluidity\cite{Eisenstein2014}. Previously, Coulomb drag study has been exclusively performed on integer $\nu_{tot}$ exciton condensate states. In GaAs double quantum wells, $\nu_{tot}=1$ is the only observed interlayer correlated state, while exciton condensation at several other integer $\nu_{tot}$ have been reported in graphene double-layer system\cite{Liu2017,Li2017}. Despite theoretical expectations\cite{Jain1989,Yoshioka1989,Scarola2001,Jain2007,Barkeshli2010,Geraedts2015}, no direct experimental observation of interlayer correlation at fractional total filling factor has been made thus far. The observed incompressible state at $\nu_{tot}=1/2$ in double-layer or wide single-layer GaAs has been proposed to be the correlated 
Halperin (331) state\cite{Halperin1983}, but without direct experimental verification\cite{Chakraborty1987,Suen1992a,Eisenstein1992a,He1992}. The delicacy of these expected interlayer FQH states demands extremely high sample quality. 

In the present study, we have fabricated monolayer graphene double-layer devices with top and bottom graphite gates. The dual graphite gates shield the graphene layers from impurities and contaminations, enabling lower disorder and more homogeneous samples\cite{Zibrov2017}. The heterostructure, stacked all at once, is composed of two graphene layers separated by hexagonal boron nitride (hBN), with the graphite/hBN encapsulation layers at the top and the bottom (Fig. 1\textbf{d}, the topmost hBN layer not shown). The thickness of interlayer hBN is $\approx$2.5~nm, which allows strong interlayer Coulomb interaction while preventing direct tunnel coupling between the graphene layers. The stack is then etched into Hall bar shape and individual contacts on each layer are fabricated. 

Coulomb drag measurements were first performed with both layers at the same carrier density ($\nu_{top}=\nu_{bot} \equiv \nu_{eq}$) (Fig. 1\textbf{a, b}). The previously observed $\nu_{tot}=1$ exciton condensate state~\cite{Liu2017,Li2017} can be clearly identified at $\nu_{eq}=1/2$, with quantized $R_{xy}^{drive}=R_{xy}^{drag}= h/e^2$ and vanishing $R_{xx}^{drag}$. In this high quality sample, however, additional features with large drag responses are also observed away from $\nu_{tot}=1$, indicating that strong interlayer coupling persists, thereby enabling additional interlayer correlated states (Fig. 1\textbf{a}). In particular, we observe vanishing $R_{xx}^{drag}$ at $\nu_{eq}= 1/4, 1/3, 2/5, 3/7, 2/3$ (data for $1/4$ is found in SI), which suggests that incompressible states are developed at these filling factors.
Among them, $\nu_{eq}=1/3$ and $2/3$ appear as trivial single layer FQH states, evident from vanishing $R_{xy}^{drag}$. We thus focus our attention first on $\nu_{eq}=2/5$ and $3/7$ particularly, which are two most prominent states that produce quantized Hall responses in the drive and drag layer. Interestingly, for these states, the two Hall resistance, $R_{xy}^{drag}$ and $R_{xy}^{drive}$, are quantized to different fractional values. For $\nu_{eq}=2/5$, we observe $R_{xy}^{drag}=1$ and $R_{xy}^{drive}=3/2$ while for $\nu_{eq}=3/7$, $R_{xy}^{drag}=2/3$ and $R_{xy}^{drive}=5/3$, respectively (from now on we use the unit of resistance quantum $h/e^2$ for the quantized resistance values). From these numbers, we note that the sum of Hall resistance in the drive and drag layer $R_{xy}^{drive}+R_{xy}^{drag}=1/\nu_{eq}$, as if a portion of Hall voltage is shifted from the drive layer to the drag layer. 

We demonstrate that $\nu_{eq}=2/5$ state can be understood with a generalized CF description extended to double-layer systems. Here, we introduce multiple species of gauge field, coupling fermions in different layers as well as in the same layer. For our purposes, we attach two intralayer flux quanta and one interlayer flux quantum to each electron, so that a CF in a given layer sees two flux quanta attached to every electron in the same layer, but only one flux quantum attached to electrons in the other layer (Fig. 1\textbf{f}). We only work in $|\nu_{top}|, |\nu_{bot}| < 1$ region, and we assume that electrons are spin and valley polarized. By generalizing the single layer CF picture, it is natural to define CF filling factors $p_A$ and $p_B$ for the top and bottom layers respectively. These are defined as the ratio between the fermion density in a given layer to the effective magnetic field felt by CFs in that layer (see SI):
\begin{equation}
\label{pnu}
p_{A} =\frac{\nu_{top} B}{B-2 n_{top} \phi_0-n_{bot} \phi_0}=\frac{\nu_{top}}{1-2 \nu_{top}-\nu_{bot}} ,   \,\,\,\,\,\, p_{B} =\frac{\nu_{bot}}{1-2 \nu_{bot}-\nu_{top}}.
\end{equation} 
Inverting Eq. \ref{pnu}, the LL filling factors for electrons in the two layers will then be given by
\begin{equation}
\label{nup}
\nu_{top} =\frac {p_A (1+p_B)}{1 + 2p_A +2p_B + 3 p_A p_B} ,   \,\,\,\,\,\, \nu_{bot} =\frac {p_B(1+p_A)}{1 + 2p_A +2p_B + 3 p_A p_B} .
\end{equation} 
In the case where the layers have equal density, this formula simplifies to $\nu_{eq} = p /(3p+1)$, where $p=p_A=p_B$.

The experimentally observed interlayer correlated state $\nu_{eq}=2/5$ corresponds to the composite fermion filling factors $p_A=p_B = -2$. Since the CFs in both layers are correlated, the Hall signal in both layers must be correlated as well. Using the CS field calculation, we find that the double layer Hall resistivity tensor obeys (see SI for derivation):
\begin{equation}
\hat{\rho}_{xy}\equiv\begin{pmatrix}
\rho_{xy}^{top}&\rho_{xy}^{drag}\\
\rho_{xy}^{drag}&\rho_{xy}^{bot}
\end{pmatrix}=\hat{\rho}_{CS}+\hat{\rho}_{cf}=\begin{pmatrix}
2&1\\
1&2
\end{pmatrix}+\begin{pmatrix}
1/p_A&0\\
0&1/p_B
\end{pmatrix}.
\label{rho}
\end{equation}
In this equation, $\hat{\rho}_{xy}$ is the Hall resistivity matrix in the unit of resistance quantum, which contains two contributing terms: $\hat{\rho}_{CS}$ originates from the motion of the CS flux considering the two intralayer flux quanta and one interlayer flux quantum, while $\hat{\rho}_{cf}$ is caused by Hall effect of CFs. At $\nu_{eq}=2/5$, Eq. (\ref{rho}) produces $R_{xy}^{drag}=1$ and $R_{xy}^{drive}=3/2$, matching the experimental observations in Fig. 1\textbf{a,b}.

Applying similar CF formalism discussed above (Eq.\ref{pnu}) to $\nu_{eq}=3/7$, however, we obtain $p_A=p_B=-3/2$, indicating that two half-filled CF LLs are involved in this state. A half-filled CF LL by itself should not develop an incompressible state. Moreover, if we were to enforce Eq. (\ref{rho}) for these values of $p_A$ and $p_B$, we would predict $\rho_{xy}^{drag}=1$ and $\rho_{xy}^{drive}=4/3$, which is in strong disagreement with the experimentally observed values, $2/3$ and $5/3$, respectively. 

In order to correct the weakly interacting CF model presented above for half-filled CF LLs, we can draw an analogy between the half-filled CF double-layer system to the half-filled electron double-layer system, in which an exciton condensate can be formed. If we assume pairing between CFs in one layer and CF holes in the second layer, 
the CF Hall resistivity tensor becomes 
\begin{equation}
\hat{\rho}_{cf}= \frac {1} {p_A+p_B} \begin{pmatrix}
1&1\\
1&1
\end{pmatrix}.
\label{eq2}
\end{equation}
Inserting Eq. (\ref{eq2}) into Eq. (\ref{rho}), we obtain $R_{xy}^{drag}=2/3$ and $R_{xy}^{drag}=5/3$, which agrees with our experimental observations,  thus suggesting the CF exciton condensation phase is indeed responsible for $\nu_{eq}=3/7$ (Further discussion of CF paring in half-filled CF LLs can be found in the SI).

Away from equal filling status, Fig. 2\textbf{a} shows that the vanishing $R_{xx}^{drag}$ persists along the segments of two symmetric lines (labeled by L1 and L2) that intersect at $\nu_{eq}=2/5$. The line L2 has a slope of -2/3 and traces from ($\nu_{top},\nu_{bot}$) = (0, 2/3) to (1, 0), while L1 is the inverse. We find that the longitudinal drag vanishes and Hall drag remain quantized along these lines, as shown in Fig. 3\textbf{a}\&\textbf{d}, indicating the strong interlayer interaction persists along these line segments. Unlike the quantized interlayer drag resistance, which is layer-independent by the Onsager relation, we find that the drive Hall resistance depends on which layer we measure. For example, along L2, we find driving the bottom layer exhibits QHE with $R_{xy}^{bot}=3/2$ and $R_{xx}^{bot}=0$. However, when we drive the top layer along L2, $R_{xx}^{top}>0$ and $R_{xy}^{top}$ is not quantized. Along L1, the role of the top and bottom layers is reversed. The experimental observed behaviors of all the resistance components along L1 and L2 are summarized in Table. 1.

We note that along L1 or L2, the composite fermion filling factor of one of the layers ($p_A$ or $p_B$) remains at $-2$ while the other can change continuously. For example, for $p_B=-2$, the filling factors $(\nu_{top},\nu_{bot})
$ given by Eq.~(\ref{nup}) satisfy the expression of L2: ($\nu_{top}+\frac{3}{2}\nu_{bot}=1$). In principle, a series of discrete incompressible FQH states can be formed along this line, corresponding to various positive and negative integer values $p_A$ in Eq.~(\ref{nup}). These should all exhibit vanishing longitudinal resistance, quantized $R_{xy}^{drag}=1$ and $R_{xy}^{bot}=3/2$, which do not depend on $p_A$, while the quantized values of $R_{xy}^{top}$ would depend on the value of $p_A$. What originally surprised us, however, is that the experimentally observed quantization of $R_{xy}^{drag}=1$ and $R_{xy}^{bot}=3/2$, together with vanishing $R_{xx}^{drag}$ and $R_{xx}^{bot}$, exists continuously along an entire segment of L2, even when $p_A$ is not an integer.

We now understand the above results as follows. For a general point on the line segment L2 (i.e. fixed $p_B=-2$), there is an energy gap for adding or removing a CF of type B ($\delta p_B$), but not of type A ($\delta p_A$). Thus, while the state should not be as stable as at a point where CF filling factors $p_A$ and $p_B$ are both integers, so that both species of CF are gapped, it should nevertheless be more stable than at a nearby points where both CF filling factors are fractions. Therefore, it is plausible that CF states along a line where one of ($p_A$, $p_B$) is an integer should be good candidates for the true ground state at the corresponding filing fractions. We call such states {\em{semi-quantized}}, as one CF filling factor is fixed but the other can vary continuously.

To understand transport properties in a semi-quantized state, we first note that in the absence of CF scattering or of pinning by impurities, there would be no longitudinal resistance and the Hall resistances would be given by Eq. (\ref{rho}), even in the absence of an energy gap. 
For the semi-quantized state along L2 ($p_B=-2$), if an electrical current is driven on the bottom layer, then the current can be carried entirely by CFs of type B (bottom drive), with stationary CFs of type A (top, drag layer). Since type B CFs are contained in a filled CF-LL, the current is carried without dissipation. Furthermore, as there is no tendency for flow of the type A CFs, a small density of impurities will have no effect, leading to quantization $\rho_{xy}^{bot} = 3/2$ and $\rho_{xy}^{drag} = 1$. On the other hand, if current is applied to the top layer, CFs of type A will be forced to move. If CFs in the partially filled CF LL are not pinned by impurities, they will participate in the motion, and the they can be scattered by impurities. This will lead to a longitudinal resistance, and deviations from the result 
$R_{xy}^{top} = (2 + 1/p_A)$ predicted by Eq.~(\ref{rho}), except in the case where $p_A$ is so close to an integer value that the small density of excess CFs is pinned by impurities.

An alternative approach to understanding properties of the states along L1 and L2 is to begin with the balanced quantized state at $\nu_{eq}=2/5$, and add quasi-holes to this state. Elementary quasi-holes in this state have total charge $-e/5$, with $-3e/5$ in one layer and $+2e/5$ in the other. The addition of one or another type of quasi-hole will move the system along the line L1 or L2, in a direction decreasing the total filling factor. The relative stability of states on the two line segments can be understood considering the energy cost for quasi-holes versus quasiparticles (See the SI for more discussion).

Finally, we turn to the state at $\nu_{eq} = 1/4 $ and the lines through it. The state $\nu_{eq} = 1/4 $ may be described in our CF language by $p_A=p_B = 1$.  As discussed in the SI, the state is also equivalent to the Halperin (331) state which has been proposed as a possible explanation for the FQH state at $\nu_{tot}=1/2$ in wide GaAs quantum wells. The line L3 in Fig. S2, which passes through this point, corresponds to $p_A = 1$ with continuously varying $p_B$.  Although there appears to be a well-developed FQH state at $p_B=2$ with $p_A = 1$ along this line, corresponding to the values $\nu_{top}=3/13, \,\, \nu_{bot} =4/13$, there does not appear to be a line of semi-quantized states between these two points.  The absence of the continuous semi-quantization along these lines suggests that the stabilization of interlayer correlated CF state requires microscopic consideration of the energetics of the quasiparticle addition to the system (see SI).

\begin{methods}
\subsection{Sample fabrication}
The hBN-graphite-hBN-graphene-hBN-graphene-hBN-graphite (from top to bottom) stack are prepared by mechanical exfoliation and van derWaals (vdW) transfer technique. The shape of graphene and graphite are carefully chosen and arranged so that we can use the overlapped part as the main channel area, while fabricating individual contacts on each layer at the regions with just one conducting layer. A square-shaped top graphene layer is chosen, while we pick strip-shaped bottom graphene and bottom graphite, that are narrower but longer than the top graphene. We align the bottom graphite and bottom graphene into a cross, while keeping the overlapped area inside the top graphene square. The top graphite covers everything after stacking but is etched into the same shape as bottom graphite. We then etch the stack into final device geometry and fabricate Cr/Pd/Au contact on all the graphene and graphite layers. Last, we grow 20-30nm ALD Al$_2$O$_3$, and then deposit contact gates above top layer graphene lead to increase the contact transparency.

\subsection{Coulomb drag measurement}

We perform Coulomb drag measurement with 2nA excitation current on the drive layer using lockin amplifiers at 17.7Hz. We used a symmetric bias scheme to eliminate any possible interlayer bias effect. In this scheme, we apply positive bias +V on the source and -V on the drain (both on the drive layer). The drag layer is open circuit but with one of the contact connected to the ground through 1MOhm resistor to allow charges to flow in and out of the layer for gating effect. Using ALD contact gate and silicon backgate, we dope the lead area of both layers to high carrier density and matching carrier type with the channel. The measurements are done in a He3 cryostat at 300mK.

\end{methods}

%% Put the bibliography here, most people will use BiBTeX in
%% which case the environment below should be replaced with
%% the \bibliography{} command.
\bibliography{Fractional_drag.bib}

\begin{thebibliography}{10}
\expandafter\ifx\csname url\endcsname\relax
  \def\url#1{\texttt{#1}}\fi
\expandafter\ifx\csname urlprefix\endcsname\relax\def\urlprefix{URL }\fi
\providecommand{\bibinfo}[2]{#2}
\providecommand{\eprint}[2][]{\url{#2}}

\bibitem{Tsui1982}
\bibinfo{author}{Tsui, D.~C.}, \bibinfo{author}{Stormer, H.~L.} \&
  \bibinfo{author}{Gossard, A.~C.}
\newblock \bibinfo{title}{{Two-dimensional magnetotransport in the extreme
  quantum limit}}.
\newblock \emph{\bibinfo{journal}{Physical Review Letters}}
  \textbf{\bibinfo{volume}{48}}, \bibinfo{pages}{1559--1562}
  (\bibinfo{year}{1982}).

\bibitem{Laughlin1983}
\bibinfo{author}{Laughlin, R.~B.}
\newblock \bibinfo{title}{{Anomalous quantum Hall effect: An incompressible
  quantum fluid with fractionally charged excitations}}.
\newblock \emph{\bibinfo{journal}{Physical Review Letters}}
  \textbf{\bibinfo{volume}{50}}, \bibinfo{pages}{1395--1398}
  (\bibinfo{year}{1983}).

\bibitem{Halperin1983}
\bibinfo{author}{Halperin, B.~I.}
\newblock \bibinfo{title}{{Theory of the quantized Hall conductance}}.
\newblock \emph{\bibinfo{journal}{Helv. Phys. Acta 56:75–104}}
  \textbf{\bibinfo{volume}{56}}, \bibinfo{pages}{75--104}
  (\bibinfo{year}{1983}).

\bibitem{Chakraborty1987}
\bibinfo{author}{Chakraborty, T.} \& \bibinfo{author}{Pietil{\"{a}}inen, P.}
\newblock \bibinfo{title}{{Fractional quantum hall effect at half-filled landau
  level in a multiple-layer electron system}}.
\newblock \emph{\bibinfo{journal}{Physical Review Letters}}
  \textbf{\bibinfo{volume}{59}}, \bibinfo{pages}{2784--2787}
  (\bibinfo{year}{1987}).

\bibitem{Eisenstein2014}
\bibinfo{author}{Eisenstein, J.}
\newblock \bibinfo{title}{{Exciton Condensation in Bilayer Quantum Hall
  Systems}}.
\newblock \emph{\bibinfo{journal}{Annual Review of Condensed Matter Physics}}
  \textbf{\bibinfo{volume}{5}}, \bibinfo{pages}{159--181}
  (\bibinfo{year}{2014}).

\bibitem{Jain2007}
\bibinfo{author}{Jain, J.}
\newblock \emph{\bibinfo{title}{{Composite fermions}}}, vol.
  \bibinfo{volume}{9780521862} (\bibinfo{publisher}{Cambridge University
  Press}, \bibinfo{address}{Cambridge}, \bibinfo{year}{2007}).
\newblock \urlprefix\url{http://ebooks.cambridge.org/ref/id/CBO9780511607561}.

\bibitem{Jain1989}
\bibinfo{author}{Jain, J.~K.}
\newblock \bibinfo{title}{{Composite-fermion approach for the fractional
  quantum Hall effect}}.
\newblock \emph{\bibinfo{journal}{Physical Review Letters}}
  \textbf{\bibinfo{volume}{63}}, \bibinfo{pages}{199--202}
  (\bibinfo{year}{1989}).

\bibitem{Halperin1984}
\bibinfo{author}{Halperin, B.~I.}
\newblock \bibinfo{title}{{Statistics of Quasiparticles and the Hierarchy of
  Fractional Quantized Hall States}}.
\newblock \emph{\bibinfo{journal}{Physical Review Letters}}
  \textbf{\bibinfo{volume}{52}}, \bibinfo{pages}{1583--1586}
  (\bibinfo{year}{1984}).

\bibitem{Suen1992a}
\bibinfo{author}{Suen, Y.~W.}, \bibinfo{author}{Engel, L.~W.},
  \bibinfo{author}{Santos, M.~B.}, \bibinfo{author}{Shayegan, M.} \&
  \bibinfo{author}{Tsui, D.~C.}
\newblock \bibinfo{title}{{Observation of a =1/2 fractional quantum Hall state
  in a double-layer electron system}}.
\newblock \emph{\bibinfo{journal}{Physical Review Letters}}
  \textbf{\bibinfo{volume}{68}}, \bibinfo{pages}{1379--1382}
  (\bibinfo{year}{1992}).

\bibitem{Eisenstein1992a}
\bibinfo{author}{Eisenstein, J.~P.}, \bibinfo{author}{Boebinger, G.~S.},
  \bibinfo{author}{Pfeiffer, L.~N.}, \bibinfo{author}{West, K.~W.} \&
  \bibinfo{author}{He, S.}
\newblock \bibinfo{title}{{New fractional quantum Hall state in double-layer
  two-dimensional electron systems}}.
\newblock \emph{\bibinfo{journal}{Physical Review Letters}}
  \textbf{\bibinfo{volume}{68}}, \bibinfo{pages}{1383--1386}
  (\bibinfo{year}{1992}).

\bibitem{Kellogg2002}
\bibinfo{author}{Kellogg, M.}, \bibinfo{author}{Spielman, I.~B.},
  \bibinfo{author}{Eisenstein, J.~P.}, \bibinfo{author}{Pfeiffer, L.~N.} \&
  \bibinfo{author}{West, K.~W.}
\newblock \bibinfo{title}{{Observation of quantized Hall drag in a strongly
  correlated bilayer electron system.}}
\newblock \emph{\bibinfo{journal}{Physical review letters}}
  \textbf{\bibinfo{volume}{88}}, \bibinfo{pages}{126804}
  (\bibinfo{year}{2002}).

\bibitem{Liu2017}
\bibinfo{author}{Liu, X.}, \bibinfo{author}{Watanabe, K.},
  \bibinfo{author}{Taniguchi, T.}, \bibinfo{author}{Halperin, B.~I.} \&
  \bibinfo{author}{Kim, P.}
\newblock \bibinfo{title}{{Quantum Hall drag of exciton condensate in
  graphene}}.
\newblock \emph{\bibinfo{journal}{Nature Physics}}
  \textbf{\bibinfo{volume}{13}}, \bibinfo{pages}{746--750}
  (\bibinfo{year}{2017}).

\bibitem{Li2017}
\bibinfo{author}{Li, J.~I.}, \bibinfo{author}{Taniguchi, T.},
  \bibinfo{author}{Watanabe, K.}, \bibinfo{author}{Hone, J.} \&
  \bibinfo{author}{Dean, C.~R.}
\newblock \bibinfo{title}{{Excitonic superfluid phase in double bilayer
  graphene}}.
\newblock \emph{\bibinfo{journal}{Nature Physics}}
  \textbf{\bibinfo{volume}{13}}, \bibinfo{pages}{751--755}
  (\bibinfo{year}{2017}).

\bibitem{Kellogg2004}
\bibinfo{author}{Kellogg, M.}, \bibinfo{author}{Eisenstein, J.~P.},
  \bibinfo{author}{Pfeiffer, L.~N.} \& \bibinfo{author}{West, K.~W.}
\newblock \bibinfo{title}{{Vanishing Hall resistance at high magnetic field in
  a double-layer two-dimensional electron system}}.
\newblock \emph{\bibinfo{journal}{Phys. Rev. Lett.}}
  \textbf{\bibinfo{volume}{93}}, \bibinfo{pages}{36801} (\bibinfo{year}{2004}).

\bibitem{Tutuc2004a}
\bibinfo{author}{Tutuc, E.}, \bibinfo{author}{Shayegan, M.} \&
  \bibinfo{author}{Huse, D.~A.}
\newblock \bibinfo{title}{{Counterflow measurements in strongly correlated GaAs
  hole bilayers: Evidence for electron-hole pairing}}.
\newblock \emph{\bibinfo{journal}{Physical Review Letters}}
  \textbf{\bibinfo{volume}{93}}, \bibinfo{pages}{036802--1}
  (\bibinfo{year}{2004}).

\bibitem{Spielman2000}
\bibinfo{author}{Spielman, I.~B.}, \bibinfo{author}{Eisenstein, J.~P.},
  \bibinfo{author}{Pfeiffer, L.~N.} \& \bibinfo{author}{West, K.~W.}
\newblock \bibinfo{title}{{Resonantly enhanced tunneling in a double layer
  quantum hall ferromagnet}}.
\newblock \emph{\bibinfo{journal}{Physical Review Letters}}
  \textbf{\bibinfo{volume}{84}}, \bibinfo{pages}{5808--5811}
  (\bibinfo{year}{2000}).

\bibitem{Yoshioka1989}
\bibinfo{author}{Yoshioka, D.}, \bibinfo{author}{MacDonald, A.~H.} \&
  \bibinfo{author}{Girvin, S.~M.}
\newblock \bibinfo{title}{{Fractional quantum Hall effect in two-layered
  systems}}.
\newblock \emph{\bibinfo{journal}{Physical Review B}}
  \textbf{\bibinfo{volume}{39}}, \bibinfo{pages}{1932--1935}
  (\bibinfo{year}{1989}).

\bibitem{Scarola2001}
\bibinfo{author}{Scarola, V.~W.} \& \bibinfo{author}{Jain, J.~K.}
\newblock \bibinfo{title}{{Phase diagram of bilayer composite fermion states}}.
\newblock \emph{\bibinfo{journal}{Physical Review B - Condensed Matter and
  Materials Physics}} \textbf{\bibinfo{volume}{64}},
  \bibinfo{pages}{853131--8531310} (\bibinfo{year}{2001}).

\bibitem{Barkeshli2010}
\bibinfo{author}{Barkeshli, M.} \& \bibinfo{author}{Wen, X.-G.}
\newblock \bibinfo{title}{{Non-Abelian two-component fractional quantum Hall
  states}}.
\newblock \emph{\bibinfo{journal}{Physical Review B}}
  \textbf{\bibinfo{volume}{82}}, \bibinfo{pages}{233301}
  (\bibinfo{year}{2010}).

\bibitem{Geraedts2015}
\bibinfo{author}{Geraedts, S.}, \bibinfo{author}{Zaletel, M.~P.},
  \bibinfo{author}{Papi{\'{c}}, Z.} \& \bibinfo{author}{Mong, R. S.~K.}
\newblock \bibinfo{title}{{Competing Abelian and non-Abelian topological orders
  in $\nu$ = 1 / 3 + 1 / 3 quantum Hall bilayers}}.
\newblock \emph{\bibinfo{journal}{Physical Review B}}
  \textbf{\bibinfo{volume}{91}}, \bibinfo{pages}{205139}
  (\bibinfo{year}{2015}).

\bibitem{He1992}
\bibinfo{author}{He, S.}, \bibinfo{author}{{Das Sarma}, S.} \&
  \bibinfo{author}{Xie, X.~C.}
\newblock \bibinfo{title}{{Quantized Hall effect and quantum phase transitions
  in coupled two-layer electron systems}}.
\newblock \emph{\bibinfo{journal}{Physical Review B}}
  \textbf{\bibinfo{volume}{47}}, \bibinfo{pages}{4394--4412}
  (\bibinfo{year}{1993}).

\bibitem{Zibrov2017}
\bibinfo{author}{Zibrov, A.~A.} \emph{et~al.}
\newblock \bibinfo{title}{{Tunable interacting composite fermion phases in a
  half-filled bilayer-graphene Landau level}}.
\newblock \emph{\bibinfo{journal}{Nature}} \textbf{\bibinfo{volume}{549}},
  \bibinfo{pages}{360--364} (\bibinfo{year}{2017}).

\end{thebibliography}

%% Here is the endmatter stuff: Supplementary Info, etc.
%% Use \item's to separate, default label is "Acknowledgements"

\begin{addendum}
 \item The major experimental work is supported by DOE (DE-SC0012260). The theoretical analysis was supported by the Science and Technology Center for Integrated Quantum Materials, NSF Grant No. DMR-1231319. P.K. acknowledges partial support from the Gordon and Betty Moore Foundation's EPiQS Initiative through Grant GBMF4543. K.W. and T.T. acknowledge supports from the Elemental Strategy
Initiative conducted by the MEXT, Japan and the CREST(JPMJCR15F3), JST. A portion of this work was performed at the National High Magnetic Field Laboratory, which is supported by the National Science Foundation Cooperative Agreement No. DMR-1157490* and the State of Florida. Nanofabrication was performed at the Center for Nanoscale Systems at Harvard, supported in part by an NSF NNIN award ECS-00335765. In preparation of this manuscript, we are aware of related work done by J.I.A Li et al. We thank Bernd Rosenow, J.I.A Li and C. Dean for helpful discussion.

\item[Author Comtributions] X.L. and P.K. conceived the experiment. X.L. and Z.H. fabricated the samples and performed the measurements. X.L. analyzed the data. B.I.H conducted the theoretical analysis. X.L., B.I.H. and P.K. wrote the paper. K.W. and T.T. supplied hBN crystals.
 
 \item[Correspondence] Correspondence and requests for materials
should be addressed to P. Kim~(email: pkim@physics.harvard.edu).
\end{addendum}

\begin{figure}
\includegraphics[width=\textwidth]{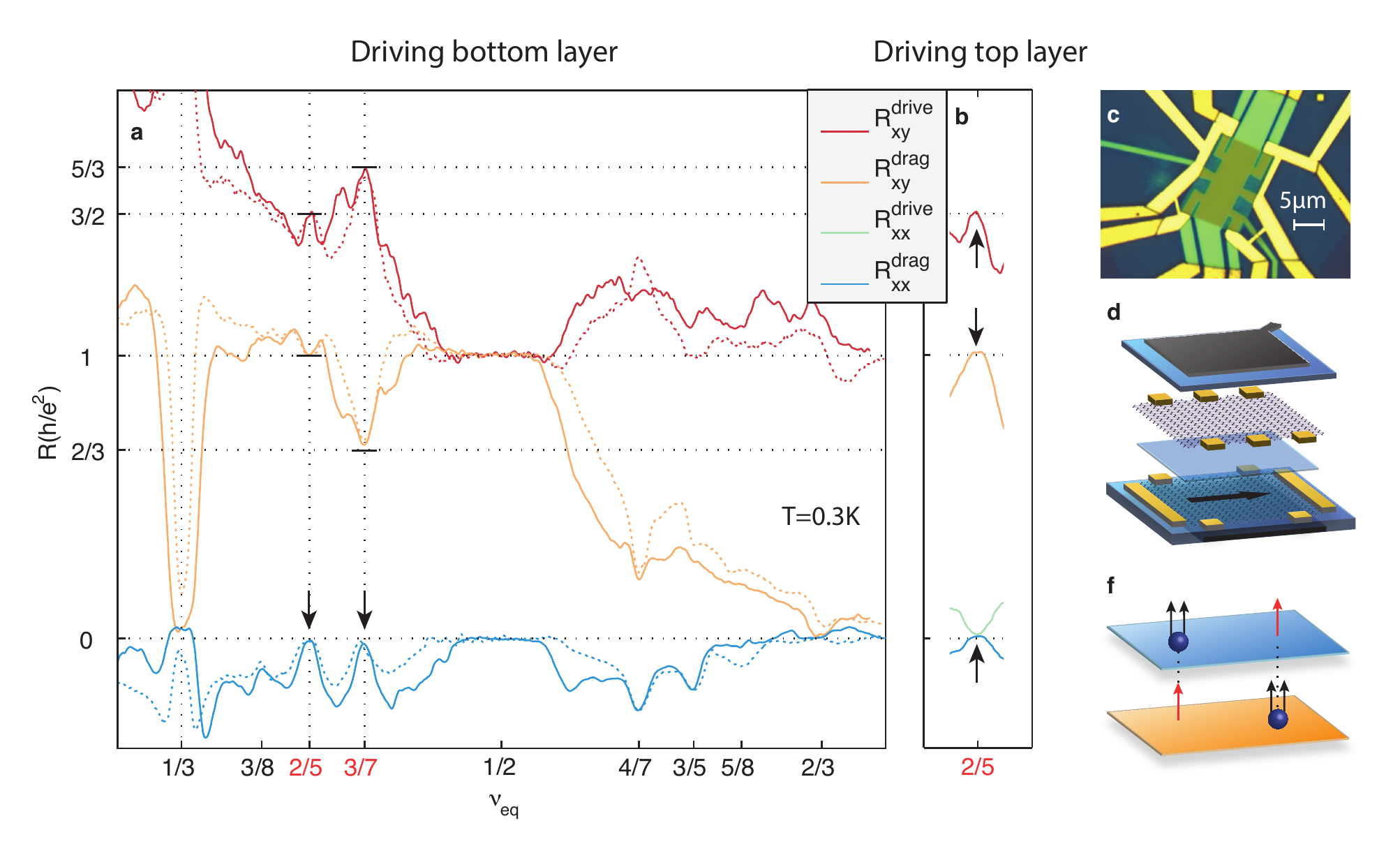}
\caption{\textbf{$\vert$ Interlayer correlated states at fractional filling factors in graphene double-layer with equal densities.} \textbf{a,} Vanishing longitudinal resistance ($R_{xx}^{drive}, R_{xx}^{drag}$) and quantized Hall resistance ($R_{xy}^{drive}, R_{xy}^{drag}$) in the drive and drag layer appear at $\nu_{eq}=\nu_{top}=\nu_{bot}=2/5$ and $3/7$. The solid curves are taken under $B=31$~T and dotted curves are from $B=25$~T. Short horizontal lines mark the Hall resistance quantization values. \textbf{b,} same measurement as \textbf{a} at 25~T, but with drive and drag layer switched. \textbf{c,} Microscope image of the device. \textbf{d,} Schematic of device structure. Graphite gates are represented by black sheets while three hBN layers are shown in blue. For this specific device, thickness of hBN in between graphene layers is 2.5~nm. Yellow blocks denote metal contacts on graphene. \textbf{f,} Illustration of Chern-Simons flux attachment. Each electron (deep blue spheres) in the system is bound with two intralayer magnetic flux quanta (black arrows) and one interlayer flux quantum (red arrows).}
\end{figure}

\begin{figure}
\includegraphics[width=\textwidth]{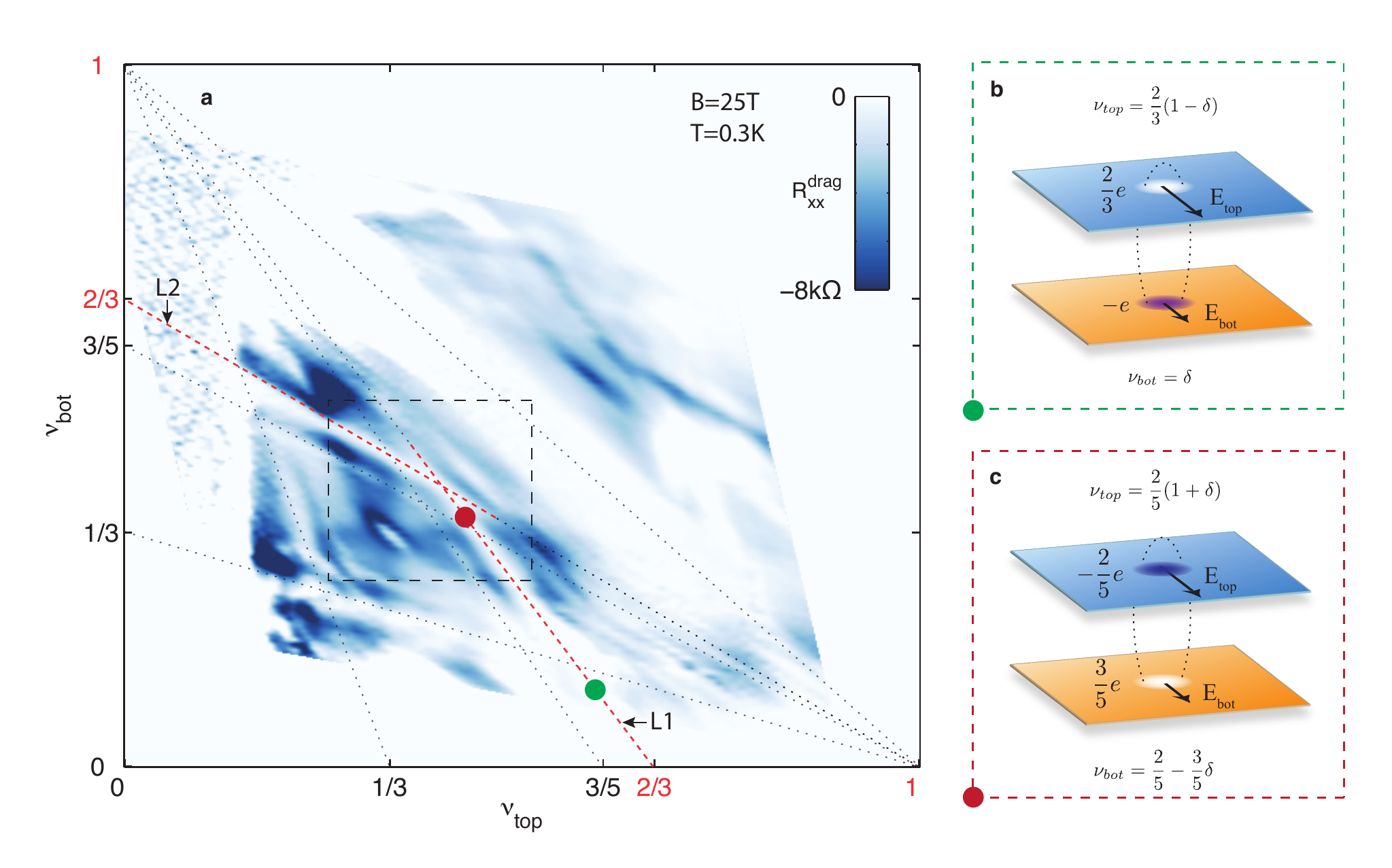}
\caption{\textbf{$\vert$ Interlayer correlation through quasiparticle pairing.} \textbf{a,} longitudinal drag resistance as a function of filing factors in the top and bottom layer. Dotted lines show locations of semi-quantized states where longitudinal drag resistance vanishes. All these lines connect filling factor $\nu=1$ in one layer with various filling factors $\nu=1/3, 3/5, 2/3, 1$ of the other layer. Among them, the intersection of red dotted lines marked by L1 and L2 corresponds to the $\nu_{eq}=2/5$ state discussed above. The dashed rectangle denotes the scope of zoomed-in measurements of Fig. 3. \textbf{b, c,} illustrations of quasiparticle pairing for two filling factor configurations (green and red dots in \textbf{a}). The circles on the two graphene layers represent quasiparticle excitations with marked electrical charges ($-e, 2/3e, etc.$). These quasiparticle pairs are balanced by the transverse electrical fields on the top and bottom layer ($E_{top}$ and $E_{bot}$ depicted by black arrows).}
\end{figure}

\begin{figure}
\includegraphics[width=\textwidth]{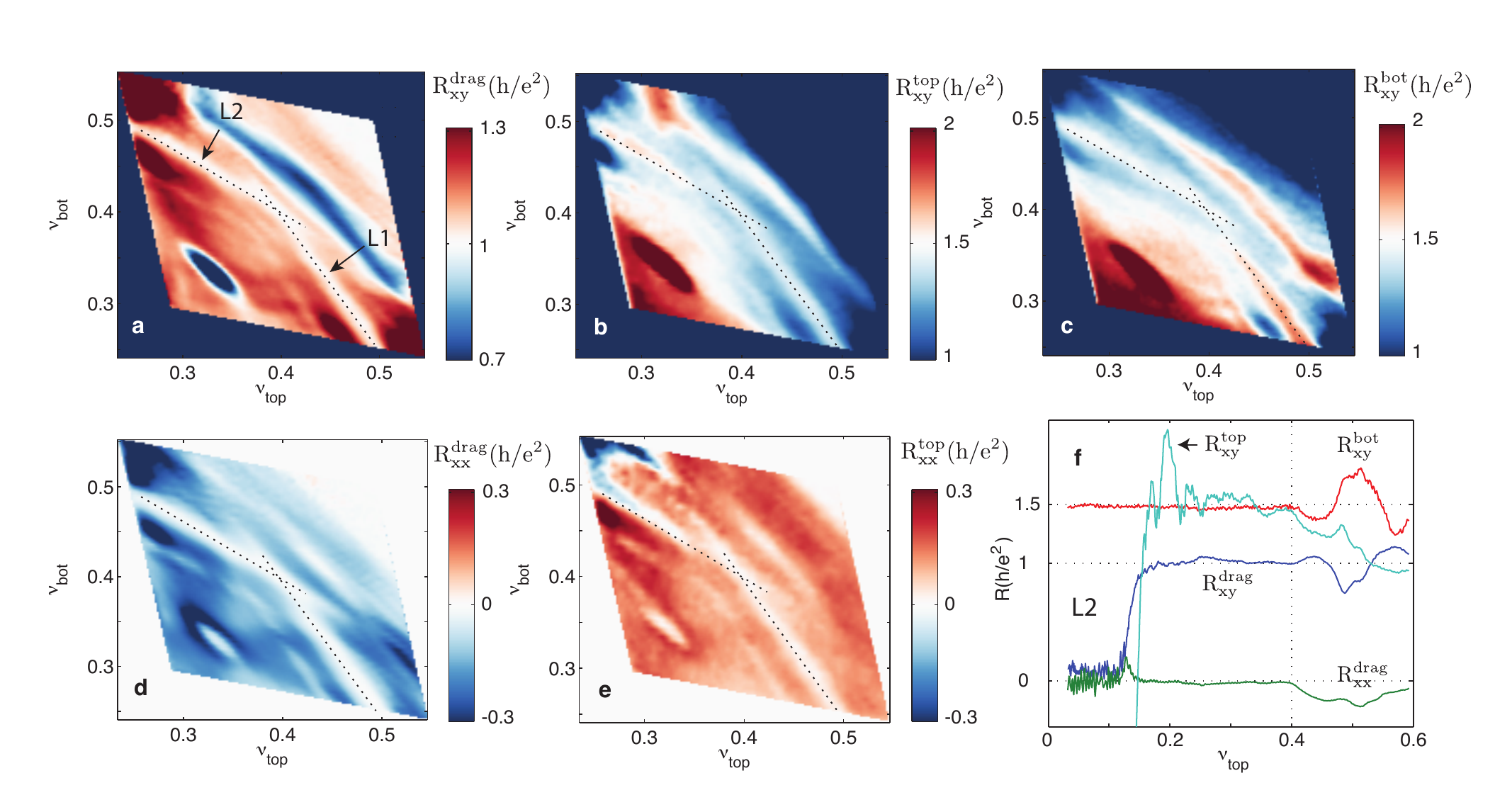}
\caption{\textbf{$\vert$ Semi-quantized fractional Hall states.} \textbf{a-e,} Various resistance measurements in the zoomed-in area indicated by dashed rectangle in Fig. 2. $R_{xx,xy}^{top}$ ($R_{xx,xy}^{bot}$) is drive layer resistance when current is driven on the top (bottom) layer. The dotted lines mark L1 and L2 (same as red lines in Fig. 2). Along L1, quantum Hall signatures ($R_{xx}^{top}=0$, $R_{xy}^{top}=3h/2e^2$) persist on the top layer but not on the bottom layer ($R_{xx}^{bot}\neq 0$, $R_{xy}^{top}\neq 3h/2e^2$) while the opposite is true for L2. Meanwhile, drag signals are quantized along both L1 and L2. \textbf{f,} Line cut through L2. It is notable that $R_{xy}^{bot}$ remained constant all the way until $\nu_{eq}=2/5$ (vertical dotted line), across the phase transition between single layer $\nu_{bot}=2/3$ fractional quantum Hall state ($R_{xy}^{drag}\approx 0$) and interlayer fractional quantum Hall state ($R_{xy}^{drag}=h/e^2$).}
\end{figure}

\begin{table}
\caption{Summary of resistance behavior along L1 and L2 shown in Fig. 2.} % title of Table
\centering % used for centering table
\vspace{2mm}
\begin{tabular}{c c c c c c c} % centered columns (4 columns)
\hline\hline %inserts double horizontal lines
& $R_{xx}^{drag}$ & $R_{xy}^{drag}$ & $R_{xx}^{top}$ & $R_{xy}^{top}$ & $R_{xx}^{bot}$ & $R_{xy}^{bot}$ \\ [0.5ex] % inserts table
%heading
\hline % inserts single horizontal line
L1 & 0 & 1 & 0 & 1.5 & $\neq$0 & non-quantized.\\ % inserting body of the table
L2 & 0 & 1 & $\neq$0 & non-quantized & 0 & 1.5 \\ [0.5ex] % [1ex] adds vertical space
\hline %inserts single line
\end{tabular}
\label{table:nonlin} % is used to refer this table in the text
\end{table}

\end{document}

\maketitle

\section*{S1. Theory of Quantized and Semi-Quantized Fractional Hall States in Coupled Layer Systems.  }
 
\subsection*{S1.1 Formalism}

Let $n_i = ( \nu_i  B / \phi_0)$ be the electron density in layer $i$, where  $i= (A,B)$ labels the top and bottom layer, respectively. Then the Chern-Simons field seen by composite fermions in layer $i$ is given by 
\be
b_i = \phi_0 ( 2 n_i + n_{i^*}),
\ee
where $i^*$ denotes the layer opposite to $i$. 
The average effective magnetic field seen by the CFs in layer $i$ may  be written as 
\be
\Delta_i = B - b_i
\ee
where $B$ is the applied magnetic field. We may then define effective Landau level filling factors for the CFs by
\be
p_i \equiv  \phi_0  n_i / \Delta_i .
\ee
We can now obtain Eq.~M1 of the main text, which expresses $\{ p_i \} $ in terms of  
$\{\nu_i\}$.

Now if voltage gradients are applied to the two layers, there will generally be electric currents generated in the layers. We consider, here, the limit of an infinite system, where the voltage gradients are produced by uniform electric fields in the respective layers, and we can neglect any contributions due to edge currents arising from differences in the chemical potentials at the two edges infinitely away. 
If  there are electron particle currents $\vj_i$ in the two layers, there will be generated Chern-Simons electric fields~\cite{Lopez1991, Halperin1993}, given by 
\be
 \vec{e}_i = \phi_0  \hat{z} \times (2 \vj_i + \vj_{i*})
\ee 
Then, If  an  electric field $\vE_i$ is applied to layer $i$, the composite fermions in layer $i$ will feel an effective electric field 
\be
\vec{F}_i = \vE_i - \vec{e}_i .
\ee
 In the limit of weak electric fields, we may use linear response and write
\be
-e   j_{i \alpha} =  \sum_{j,\beta} \hat{\sigma}^{cf}_{i  \alpha, j \beta} F_{j \beta} ,
\ee
where $\alpha$ \& $\beta$ are indexes of two spacial coordinates and $\hat{\sigma}^{cf}$ is $4 \times 4$ matrix that we denote as the conductivity matrix for CFs.  If we now invert these equations to express the electric fields in terms of the electric currents, we find a resistivity matrix given by 
\be
\hat{\rho} = \hat{\rho}^{CS} + \hat{\rho}^{cf}  ,
\ee
where $\hat{\rho}^{cf} $ is the matrix inverse of $\hat{\sigma}^{cf} $ and
\be
\rho^{CS}_{ i \alpha, j \beta} =   \epsilon_{\alpha \beta} ( 1 + \delta_{ij})  \, (h/e^2)   .
\ee

In a self-consistent  mean field or random phase approximation,  one assumes that the response function $\hat{\sigma}^{cf}$
is given by the response function of free fermions to the applied forces $\vec{F}_i$ in the effective magnetic   fields $\Delta_i$.  In this case,  $\hat{\rho}^{cf}$ , has no elements connecting the two layers. In addition, if we are in a situation were there is no dissipation, the longitudinal  part of $\hat{\rho}^{cf}$ will be zero, and only the Hall components survive. Moreover, in the absence of pinning by disorder,  the CF Hall resistance of layer $i$ is just equal to $1 / (p_i \phi_0)$ .  The $2 \times 2$ matrix of Hall resistances is then given by Eq.~(M3) of the main text, in units where the resistance quantum $h/e^2$ has been set equal to unity.

\subsection* {S1.2 Quantized  and Semi-Quantized Fractional Hall States}

Fully quantized fractional Hall states can arise if both $p_A$ and $p_B$ are integers.  Then, at least within mean field theory, there will be an energy gap for creation of any  kind of excitation, and the state should be stable against small changes in the chemical potential of either layer. (Whether the proposed state is  actually the true ground state of the system, with a lower energy than any other possible state, will of course depend on details of the Hamiltonian).  When there  is  a complete energy gap,  there will be effectively no excited  quasiparticles at sufficiently low temperatures in any of the composite-fermion Landau levels, so there will  be no dissipation, and the resistivity matrix will be given by Eq.~(M3). 

However, we may also consider a situation where only one of the filling factors, say $p_B$ is an integer.  This means that there should be energy gap for changing $p_B$ but not for $p_A$:  the ground state energy should vary continuously with $p_A$, if $p_B$ is held fixed.  
With some simple algebra on Eq. (M2), one can show that the resulting filling factors lie on a straight line, with 
\be
\nu_A + \alpha \nu_B= 1 \, , \,\,\,\,\,\,\, \alpha = \frac {2 p_B + 1}{p_B} \, . 
\ee
We refer to generic states along these lines, at points where only one of the occupation numbers $p_A,p_B$ is integer, as 
 semi-quantized fractional Hall states.

As argued in the main text, if the drive current is applied to layer B, there will be no dissipation. In the present language, this is because $\vec{F}_A = 0$ and the CFs of type B reside in CF-LLs that are completely full.   Consequently, 
 we can still use Eq.~(M3) to compute the Hall resistivities.  Thus $\hat {\rho}^B_{xy}$ and  $\hat {\rho}^{drag}_{xy}$ should still remain quantized, with values  $ (2 + 1/p_B$ ) and 1, respectively, in units of $h/e^2$.   However, if drive current is applied to layer A, then  $\vec{F}_A $ will be nonzero, and there will generally be dissipation due to scattering of the CFs by impurities,  Therefore, the prediction of Eq.~(M3) that 
$\hat {\rho}^B_{xy} = (2+1/p_A)$  cannot be trusted. However, we note that if 
$p_A$ is sufficiently close to an integer value, and the temperature is sufficiently low, the composite fermion  particles or holes in the partially filled Landau level may be localized by disorder, so that they will  not contribute to the transport properties. In this case, one should replace the quantity $p_A$ in Eq.~(M3)  by the nearby integer value.

In contrast to the semi-quantized case, we may consider the unquantized case, where both $p_A$ and $p_B$ are non-integer.  In this case we have partially filled Landau levels for composite fermions of both species.  Now, even in the absence of disorder, composite fermions of the two types will scatter off each other, unless their drift velocities are identical. The velocities will only be equal if the electrical field is identical in the two layers, which is not the case in the drag experiments we are considering.  Thus we would expect to find dissipation even in the absence of impurities.

For unquantized states along the symmetric line $\nu_A=\nu_B$,  if we were apply equal current densities in the two layers, the electric fields would also be equal in the two layers. Therefore, in the absence of disorder, there should be no dissipation in this case, and there will be no longitudinal voltage drop in either layer.   On the other hand, if the currents are equal and opposite in the two layers,  the electric fields should be opposite in the two layers, and there would be dissipation due to the scattering between composite fermions in opposite layers, if $p_A$ and $p_B$ both not integers. In this case,  in the absence of impurities, we  will have nonzero  longitudinal electric fields which are equal and opposite in the two layers. 

The longitudinal voltage in a drag experiment, where current flows only in one layer, may be obtained by adding the two previous cases.  We see that  
in the absence of disorder, the longitudinal electric field should remain equal and opposite in the two layers,
 {\it {i.e.}}, for the unquantized states along the symmetry line, we should have  $ \rho^{drag}_{xx} = - \rho^{drive}_{xx}.$ 
  Experimental results along the symmetry line are at least approximately  consistent with this expectation, suggesting that  scattering between the two types of composite fermions is more important than the scattering by disorder in this case. 

\subsection*{S1.3 Quasiparticles} 

For a fully quantized system, with a proper energy gap, the elementary excitations are  found to be quasiparticles and quasi-holes,   with fractional charges that are precisely quantized in each of the two layers.
As an example, we may consider the gapped symmetric case at $(2/5, \, 2/5)$.  Generalizing  the Laughlin procedure~\cite{Laughlin1983}, we may generate a quasiparticle  by inserting a zero-diameter solenoid at a point $\vr_0$, and turning on current in the solenoid  so that it contains precisely one quantum of magnetic flux.  Since we have a symmetric system with total filling $\nu_{tot} = 4/5$, this procedure will expel 2/5 of an electron charge from the vicinity of $\vr_0$ in each layer, and send it to the boundary of the sample.  If we now deposit an extra electron at point $\vr_0$ in layer $A$, we will have created an ``A-type"   quasiparticle, with 3/5 of an electron charge in layer $A$, and -2/5 in layer $B$, for a total charge of -e/5.  We may create an ``A-type"   quasi-hole,  with -3/5 of an electron charge in layer $A$, and 2/5 in layer $B$,  using the reverse procedure, where we change the sign of the flux through the solenoid and remove an electron from layer A.  Type-B  quasiparticles or quasi-holes can be created in the same way, by adding or subtracting the electron in layer B.  By adding or subtracting  quasiparticles or quasi-holes of type A,  we can shift the filling factors $(\nu_A, \nu_B)$ along the line L2, with slope -2/3, whereas by adding or subtracting quasiparticles of type B, we move along the line L1, with slope -3/2.

If the density of quasiparticles is sufficiently small, so that they do not overlap, we can continue to talk about quasiparticles  with quantized charges.  At some point, however, the charge associated with a particular quasiparticle becomes ambiguous, and only the total charge in each layer is well defined. If one progresses far enough  from the (2/5,2/5) state along the line L2 with slope -2/3, one will pass through a number of states with $p_B = -2$ and integer $p_A$.  If our CF model is still a correct description of the system, then the ground states should be fully quantized states, with complete energy gaps.  Then, if the temperature is  well below the energy gap, the charged excitations will be quasiparticles and quasi-holes, with quantized charges, which will generally not be the same as those at (2/5, 2/5). However, in each case, we can find quasiparticles with minimal total charge and with a ratio of charges of -2/3 between layer B and layer A.  For example, at the end point (0,2/3) we would find quasiparticles with charge 1 in layer $A$ and -2/3 in layer $B$.  (We may understand this quasiparticle as a bound state of an electron in the empty layer with a pair of $e/3$ quasi-holes in the  layer with $\nu_i=2/3$.) In fact, experimentally, it appears that the CF model where each CF sees  two  flux quanta attached to CFs in the same  layer and one for the opposite layer is not the correct description for fractions with $\nu_A \leq \, \approx 0.2$,  but the example illustrates the point. More generally, for fully quantized states located on the the line, we would find quasiparticles whose charges in the two layers are given by
\be
q_A = 3 q_T  \, , \,\,\,\,\, q_B = -2q_T \, , \,\,\,\,\,\,\, q_T = \frac {p_A+ 1}{4 p_A + 3} \, ,
\ee
with $q_T$ being the total charge of the quasiparticle. 
Thus in all these cases, one could move along the semi-quantized line by adding quasiparticles of the designated type.  

Transport properties of a semi-quantized state can be understood in terms of the properties of quasipaprticles, without referring to the motion of CFs.  At a point on L2 with $\nu_{top}$ slightly less than 2/5,  we may describe the ground state as the quantized (2/5,2/5) state, with a nonzero  density of type-A quasi-holes.   If a current is sent through the bottom layer (layer B), the current can be carried by the quantized background state, which as we have seen will generate a Hall voltage in layer A that is  2/3 the Hall voltage in  layer B.  Since the type-A quasi-holes have a charge ratio of -3/2, the electric fields in the two layers exert no net force on the quasi-holes.  Also,  current carried by the background state engenders no longitudinal electric field in either layer. Consequently,  it is consistent that the quasiparticles will not move and there is no diissipation.  
By contrast, if the current is applied to layer A,  the  Hall field in layer A will be 3/2  times the field in layer B, so there will be a net force on the quasi-holes.  If the quasi-holes are not pinned, they will move under this force and dissipation will result.

 \subsection*{S1.4 Model for the (3/7, 3/7) State}
 
 The discussion in the previous sections has been concerned with states where at least one of the parameters $p_A,p_B$ is an integer, and the ground state can be understood using a mean field theory of non-interacting CFs. This is not the case for the quantized Hall state at  (3/7, 3/7), where $p_A=p_B= -3/2$.
Now  we have one full effective Landau level, and one half-full level for each species of CF.  For the half-full level, we must take into account interactions between the CFs. We shall assume that there is a pairing between particles of one species and holes in the other, so that the wave function for composite fermions in the partly-filled level can be described by the familiar (111) state.  Alternatively, we may say that there is a superfluid Bose condensate of excitons formed from the composite fermions.  In this state, there can be no difference in the effective electric field $\vec{F}_i$ felt by the CFs in different layers, regardless of any difference in the currents between layers.  Moreover, if current is distributed so that the CFs have the same velocity in the two layers,
then  interactions between the CFs play no role, and the effective field should have the same value as for uncoupled layers, 
$\vec{F}_i = \hat{z} \times \vec{j} / (p_A+p_B)$, where $\vec{j}$ is the total current density int the two layers.
This implies that the Hall resistivity matrix for CFs is given by Eq.~(M4) of the main text, which is in agreement with our experimental observations. 

We may contrast these results with what one might have found from some alternate models of the (3/7, 3/7) state.
For example, one could have supposed that the quantization resulted because CFs in the half-full LLs form Cooper pairs  separately within each layer, in analogy with the models that have been proposed to explain the FQH state at $\nu=5/2$ in GaAs~\cite{Read2000}.  This would result in a diagonal form for the CF Hall resistivity matrix, identical to that in Eq.~(M3) of the main text. Alternatively, one could have imagined that Cooper pairs are formed between CFs in opposite layers (as opposed to pairing between CFs and holes). This would result in a CF Hall resistance matrix proportional to 
$  \begin{pmatrix}
1& -1\\
-1&1
\end{pmatrix} $.
In either of these cases, the results predicted for the Hall drag resistivity would disagree with our experimental results.

  \subsection*{S1.5 Energy gap and incompressibility}
  
  We have found, in the case of half-integer $p$, that the mean-field description of composite fermions reduces to a situation in which an effective  Landau level has two species of fermions with total filling equal to unity.  We have assumed that interactions are such that the ground state for the fermions has a Bose condensate of exciton pairs, which means that the ground state has a bosonic order parameter, characterized by an arbitrary phase $\phi$.  If one were to neglect coupling to  the Chern-Simons gauge fields, one would expect that the state would be compressible, and that it would have a Goldstone mode whose energy would vanish in the limit of long wavelengths.  However, we argue here that coupling to the gauge fields will cause the actual system to be incompressible and will therefore lead to an energy gap.
  
  In particular, consider a long-wavelength fluctuation in which there is, locally, a difference in the  electron density of the two layers, while the total density is held constant.  This will give rise to a difference in the effective magnetic fields, $\Delta_A$ and $\Delta_B$ felt by the composite fermion species, and consequently a difference in the vector potentials felt by the two species. 
From gauge invariance, we would expect a term in the energy density of the form 
\be
\epsilon \propto \rho_s  ( \nabla \phi - \vec{a})^2 , 
\ee
 where $\rho_s$ is a stiffness constant,   and  $\vec{a}$ is the difference in vector potentials for the two species.  The coupling of the phase gradient and the vector potential is determined by the fact that $\phi$ is canonically conjugate to fluctuations in the exciton density, which is itself the difference in densities of fermions in the two layers. Now if the density difference has a form $\delta n = c \cos qx$ ($c$ is a constant), there will be a vector potential $\vec{a}$ in the $y$-direction with a magnitude $\left| \vec{a} \right| \propto (c/q)  \sin qx$.  Since the density does not depend on $y$, one will not find a $y$-component of the phase gradient, at least  in linear response, and thus there will be an  energy cost proportional to $(c/q)^2.$ 
 
For any fixed value of the wave vector $q$, if the density fluctuation is sufficiently large, it will become favorable to introduce vortices in the bosonic order parameter, and thereby reduce the energy cost below $(c/q)^2$.  In the long-wavelength limit, this will occur even for small density fluctuations, and the density of vortices will be proportional to the size of the density fluctuation.  Thus the actual energy cost will be linear in the magnitude of the deviation from equal filling, and we say that the state is incompressible.

We note that  an isolated vortex in this state will have a finite energy.  In the the absence of coupling to the gauge fields, a vortex would have a logarithmically diverging energy.  However, coupling to the gauge fields allows for screening, so that far from the vortex the gradient in $\phi$ is canceled by the resulting vector potential $\vec{a}$.  This requires that there be a charge accumulation near the vortex core, which will be equal and opposite in the two layers.  For a vortex of unit circulation,  the charge in each layer will be $\pm 1/(6p_A+2)$, which,  in the case of the (3/7, 3/7) state, is  $\pm1/7$ of an electron charge.

\subsection*{S1.6 Deviations from (3/7, 3/7)}

Small deviations from the stable point (3/7, 3,7) along the line where $\nu_{tot} = 6/7$ can be achieved by adding vortex excitations, which have total charge zero but non-zero charge in each layer separately.  Although there will be an energy cost for adding these vortices, it is plausible that the energy cost may be smaller than the energy to add  a quasiparticle with net charge different from zero.  In this case, one should find a valley of relatively stable configurations along the line $\nu_{total} = 6/7$,  in the vicinity of  (3/7, 3/7).  These states are similar to those in the valleys emanating from quantized states with integer $p$ in the sense that they are achieved by adding to the parent state quasiparticles of a single type, which have a fixed ratio of the charge in each layer.  Therefore, if the ratio of the electric fields in the two layers is properly chosen, the quasiparticles will not feel a force and can  remain stationary, even in the absence of pinning, and there will be no dissipation. 

In the present case, the absence of dissipation should  occur if the electric fields are equal in the two layers.   If the electric fields unequal, the quasiparticles will move, which will lead to dissipation if there is scattering due to impurities.  However, the dissipation would still vanish in the absence of impurities, as there is only one type of quasiparticle, and they would all drift at the same velocity. 

We may note that for small deviations from the symmetric point, states along the line $\nu_{total} = 6/7$  can be parametrized by $(p_A = -3/2 + \epsilon , \,\, p_B= -3/2 - \epsilon)$.  However, for larger values of $\epsilon$, the resulting filling factors do not lie on this line but rather on a curve which passes through the points (1,0) and (0,1). In any case, once the deviation from (3/7, 3/7) becomes large, interactions between quasiparticles will be important, and we cannot say much about the resulting state. 

In our experiments, we see some evidence for semi-quantized states along a line of slope -1 in the vicinity of  the point (3/7, 3/7).  However, these states do not extend very far from symmetric filling, and we do not observe evidence for curvature of the line. 

\subsection*{S1.7 Relation to trial wavefunctions}

The mean field ground states that we have  described in the language of fermions coupled to Chern-Simons  fields, have direct counterparts in the language of CF trial wave functions, introduced by Jain~\cite{Jain2007}. The trial wave function corresponding to a state with CF fillings $(  p_A, \, p_B)$ can be written as 
\be
\Psi=  P_{LLL} \Psi_{p_A, \, p_B } \{ z_i, w_k \}  \prod_{i<j} (z_i - z_j)^2   \prod_{k<l} (w_k - w_l)^2      \prod_{i,k} (z_i - w_k),
\ee
where $z_i$ and $w_k$ are the positions of electrons in the two layers, in complex coordinates, $\Psi_{p_A, \, p_B } \{ z_i, w_k \}$ is the wave functions for a set of non-interacting electrons in a state with $p_A$ filled LLs in layer A and $p_B$ filled LLs in layer B, the operator $P_{LLL}$ represents projection to the lowest LL,  and we have omitted the single-particle Gaussian factors. 

In the case of our quantized Hall state at (1/4, \,1/4), corresponding to $p_A=p_B= 1$, the wave function $ \Psi_{p_A, \, p_B } $ is already confined to the lowest LL, and the projection operator can be omitted. The wave function in this case may be written as 
\be
\Psi_{331} =  \prod_{i<j} (z_i - z_j)^3   \prod_{k<l} (w_k - w_l)^3      \prod_{i,k} (z_i - w_k) , 
 \ee
which is identical to the  Halperin (331) wave function for an FQH state with $\nu_{tot}=1/2$, which is  considered a possible model for such observed states in wide GaAs quantum wells.

Another interesting case is  generalization of the (3/7,3/7) state to a state with $p_A = p_B = 1/2$.  If the function   $ \Psi_{p_A, \, p_B } $ is chosen to be the Halperin (111) wave function, which is a description of the interlayer coherent IQH state with equal filling in the two layers,  then the final  wave function takes the form:
\be
 \Psi_{332} =  \prod_{i<j} (z_i - z_j)^3   \prod_{k<l} (w_k - w_l)^3      \prod_{i,k} (z_i - w_k)^2 ,.
 \ee
 This is just the Halperin (332) wave function which was proposed for the ground state  at $\nu_{tot} = 2/5$,  for a collection of spin-1/2 electrons with negligible Zeeman splitting.

\subsection*{S1.8 Fractions with $1 < \nu_{tot } <  2$ }

Our discussion so far has been concerned with the region of the phase diagram with $ 0< \nu_{tot} < 1$. We have assumed that in this region, the partially filled Landau level consists of electrons in a single valley and a single spin direction. It is likely that 
in the region $1 < \nu_{tot } <  2$, the partially filled Landau level again contains electrons in a single valley and spin state, with either the valley or the spin state or both reversed from the region $\nu_{tot}<1$.  If this assumption is correct,  states in the two regions should be related by a particle-hole transformation. Specifically, if we find a quantized state  $(\nu_{top} , \nu_{bot})$,  there should be another quantized state $(\nu'_{top} , \nu'_{bot})$, with $\nu'_i = 1 - \nu_i$. The corresponding Hall matrices should be related by 
 \be
 \hat{\rho}' = [ 1 - (\hat{\rho})^{-1} ] ^ {-1} .
 \ee

\subsection*{S1.9 Systems of finite size} 

Our analysis so far has focused on situation of an infinite system, where transport takes pace in the bulk, and the contribution of edge states can be neglected. For a quantized Hall state in a finite system, there will be contributions from both the edge and the bulk.  It is convenient to break the electrochemical potential at any point into a part that arises from the macroscopic electrostatic potential and a chemical potential,  arising from the electron kinetic energy and the microscopic atomic potentials, which depends largely on the local electron density and magnetic field. If there is a difference in electrostatic potential between opposite edges of a sample, there will be electric fields in the bulk, which lead to bulk Hall currents, as we have calculated above. If there is also a  chemical potential difference between  the edges, there will be edge currents due to the difference in occupations of current-carrying states at the two edges. However, it is well established that the total Hall current in a quantized state is determined only by the voltage difference between opposite edge states and is independent of the division between electrostatic and chemical potential~\cite{Halperin1982}.

If we now consider a semi-quantized state along the line L2, and apply the drive current to the bottom layer, where $p_B = -2$, and we neglect any possible motion of the quasiparticles of type A, then we will obtain the same voltage differences between opposite edges for a finite system, for a given total current,  as in the limit of an infinite system.  Motion of the quasiparticles will be driven by gradients in the chemical potential (diffusion) as well as by the gradient of the electrostatic potential. In our situation,  the total Hall voltage in the top layer will be  3/2 the Hall voltage in the bottom layer,  while the  type-A quasiparticles have charges of opposite sign in two layers, with a ratio -2/3,. Thus,  we would expect that on average there will be no driving force for motion of  the quasiparticles, and it is consistent to assume they do not contribute to the current.   By contrast, if the drive current is applied  to the top layer, there will be a net force on the quasiparticles, and dissipative motion will result.  Similar considerations show that our predictions for the drag and drive resistances in semi-quantized states should apply to finite systems as well as in the limit of an infinite system. 

\subsection*{S1.10 Simplified model of transport in a semi-quantized state}

We present here a less rigorous, but perhaps more intuitive way to understand the Hall resistance quantization along L1 and L2. For example, let us model states along L2 by starting from the end-point (0, 2/3),  considering filling factors $(\nu_{top}, \nu_{bot})=[\delta, 2/3(1-\delta)] $, with $\delta$ small.  (Fig. S1). The state (0, 2/3) consists of a completely empty top layer and a bottom layer in a conventional quantized Hall state with $\nu = 2/3.$  We may suppose that for a suitable strength of interaction between the layers, an electron added to the top layer would bind strongly to a pair of $e/3 $ quasi-holes in the top layer, giving rise to a combined quasiparticle whose total charge is one-third of an electron charge, $-e/3$.  Further, we may suppose that the state with $\delta \neq 0$ is formed by addition of a finite density of these compound quasiparticles to the  (0,2/3) ground state.  .

If a current $I$ is driven on the bottom layer, it will be carried solely by electrons in the $\nu=2/3$ ground state, with no motion of the quasiparticles.  Schematically, we can say that the current is carried by the FQH edge state of the  $\nu=2/3$ layer. (Fig. S1{a}).  The quasiparticles cannot contribute a net current, because that would require a current of electrons in the top layer, in violation of the boundary condition in the drag geometry. The absence of quasiparticle motion requires, in turn, that the net force on a quasiparticle is zero, which will occur if and only if the electric field in the lower layer is 3/2 times the field in the bottom layer.  Furthermore, in order to get a net current $I$ from the edge states in the lower layer, the Hall voltage in that layer must be equal to $3I h/2 e^2$.  Thus we find $R^{drive}_{xy} = 3/2, \, R^{drag}_{xy}= 1$, in units of $h/e^2$. 

In the case where we drive current $I$ on the top layer, the current must be carried by quasiparticles, since there are no edge states in the upper layer.  (Fig. S1(b)). This current drags a quasiparticle current of $-2I/3$, in the opposite direction, on the bottom layer. This current must be canceled by  a $2I/3$ edge current on the bottom layer, which requires that there be a Hall voltage  $V_{xy}^{drag}=\frac{3h}{2e^2}\times\frac{2}{3}I=\frac{h}{e^2}I$. 
By contrast, the voltage  $V_{xx}$ and $V_{xy}$ in the top layer will not be quantized, and will depend on details such as the amount of impurity scattering, as well as the value of $\delta$.
Thus we have a quantized Hall drag effect in this configuration, despite having no quantized Hall effect in the top layer. 

\begin{figure*}
\centering
\includegraphics[scale=0.9]{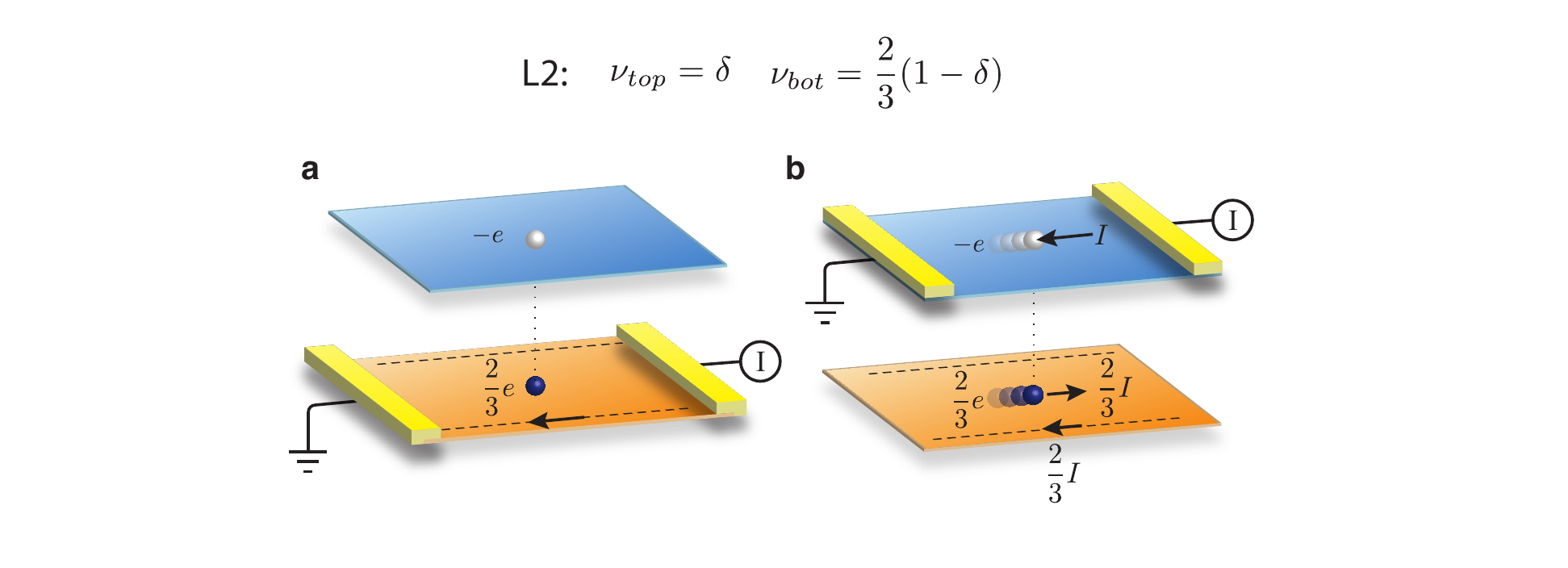}
\caption{({\bf{a, b}}), Schematic of transport in a semi-quantized state along the  line L2, starting from the (0, 2/3) reference state.  The dashed line on the edge of the bottom layer represent a 2/3 FQH edge state.  Quasiparticles in this model consist of an electron in the nearly-empty top layer (white sphere) bound to a pair of anyons (quasi-holes) with total charge $2e/3$ in the bottom layer. 
({\bf{a}}).   When current is driven in the bottom layer,  all the current can be carried by the 2/3 edge state (black arrow), while the quasiparticles remain stationary.  ({\bf{b}}). Current flowing on the top layer is carried solely by the quasiparticles, due to the absence of an edge state on the top layer.  Hence, a drive current $I$ in the top layer drags along a current $-2I/3$ in the bottom layer, which must be canceled by an edge current $2I/3$ in the opposite direction. This requires that there be a Hall voltage in the bottom layer, $V^{drag}_{xy} = (2I/3) (3h/2e^2) = I (h/e^2) $.  Thus we have quantized drag in this configuration, despite no quantized Hall effect in the top layer.}{\label{FS1}}
\end{figure*}

\section*{S2. Additional data and analysis} 
\subsection*{S2.1 Line cuts through (1/4, 1/4) and (2/5, 2/5) state}
In Fig. S2, we can see at $(\nu_{top}, \nu_{bot})=(1/4, 1/4)$, the Hall drag is quantized to $h/e^2$ while the Hall resistance of the drive layer approaches $3h/e^2$. The longitudinal drag resistance also vanishes. Using the Composite Fermion transformation in main text Eq.~(M1), we found that the (1/4, 1/4) state corresponds to $(p_{B}, p_{B})=(1, 1)$. And applying Eq.~(M3), we found indeed that drive layer Hall resistance should be $3h/e^2$ while Hall drag is $h/e^2$, matching experimental values. 

Proceeding along L3 away from $(\nu_{top}, \nu_{bot})=(1/4, 1/4)$ (Fig. S2), this line cut reveals the quantized Hall states exist for integer composite filling factors $(p_A, p_B)=(1, 1)$ or $(1, 2)$. However in between $(p_A, p_B)=(1, 1)$ and $(1, 2)$, Hall-drag and longitudinal drag do not stay quantized. The lack of semi-quantized Hall state behavior is interesting and are different from the case of line cut through $(\nu_{top}, \nu_{bot})=(2/5, 2/5)$ (Fig. 3f).  This suggest that Eq.~(M3) cannot be blindly applied at arbitrary $(p_A, p_B)$ state, even when one of $p_A$ or $p_B$ is an integer.

Although our discussions are focused on line cuts along L2 for (2/5,2/5) and L3 for (1/4, 1/4), similar conclusions can be reached from line cuts following L1 and L4. Along L1 (Fig. S3), semi-quantized state is observed, similar to the case of L2 but with the top layer as the quantized layer. Following L4 (Fig. S2), Hall-drag and longitudinal drag is not quantized in general except for $(p_A, p_B)=(1, 1)$ and $(2, 1)$.

\begin{figure*}
 \centering
 \includegraphics[scale=0.8]{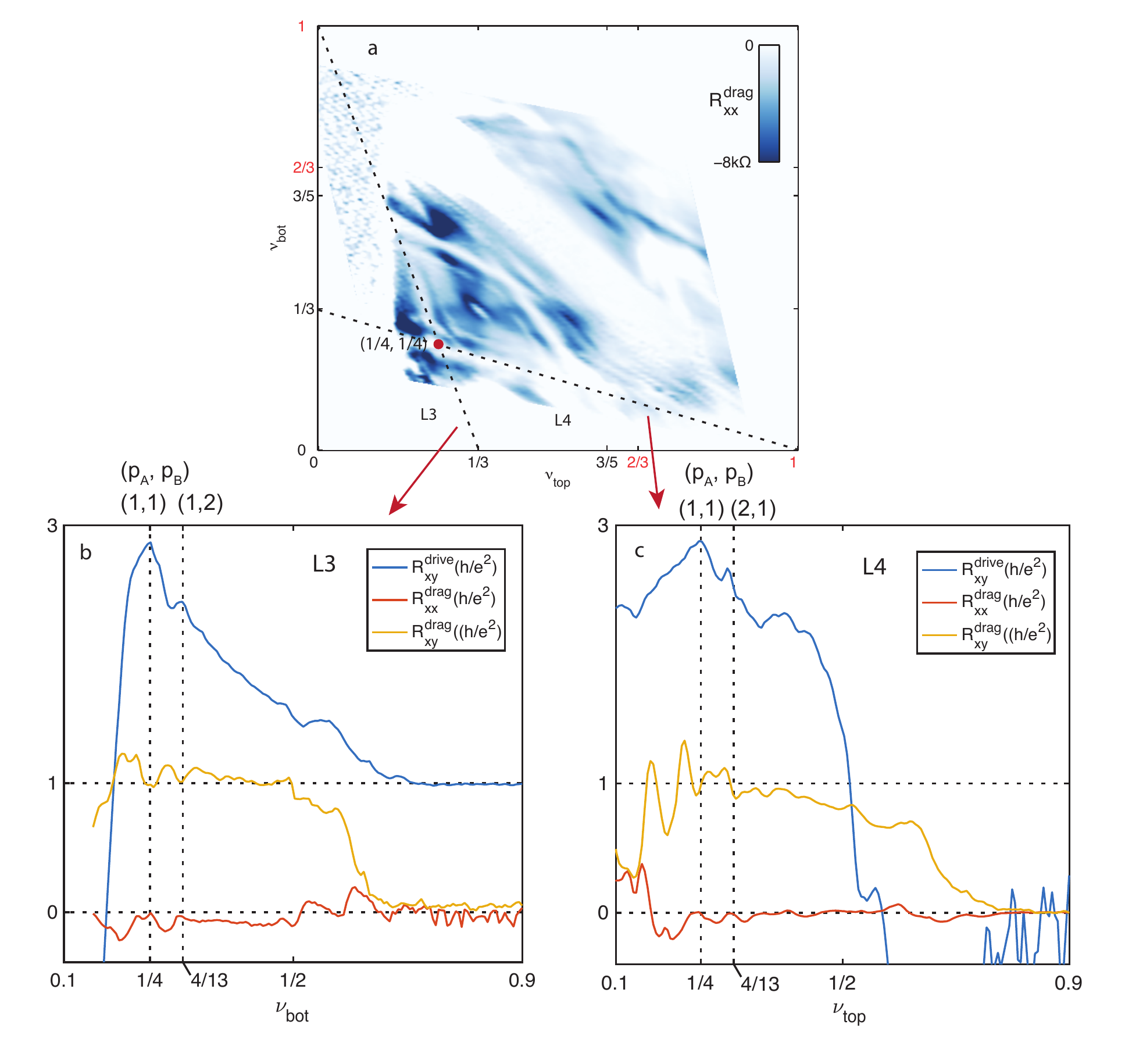}
 \caption{(a) map of longitudinal drag resistance showing where the line cuts are being taken. (b) linecut L3 through $(\nu_{top}, \nu_{bot})=(1/4, 1/4)$. At $\nu_{top}=\nu_{bot}=1/4$ (left vertical dashed line), the longitudinal drag vanish and Hall drag quantize to $h/e^2$. At the same time, the Hall resistance of the drive layer is close to the expected quantization value of $3h/e^2$. After composite fermion tranformation, $\nu_{top}=\nu_{bot}=1/4$ becomes $(p_A, p_B)=(1, 1)$, corresponding to a quantized interlayer state. Going away from $\nu_{top}=\nu_{bot}=1/4$ along this linecut, the quantization is lost. However Hall drag and magneto-drag quantization recovers near $(\nu_{top}, \nu_{bot})=(3/13, 4/13)$ (right vertical dashed line). This filling factor correspond to composite fermion filling $(p_A, p_B)=(1, 2)$, another quantized interlayer state. (c), linecut L4 through $(\nu_{top}, \nu_{bot})=(1/4, 1/4)$ in a different direction. The data quality is worse than (b), but show the same behavior. }{\label{FS2}}

\end{figure*}

\begin{figure*}
 \centering
\includegraphics[scale=0.8]{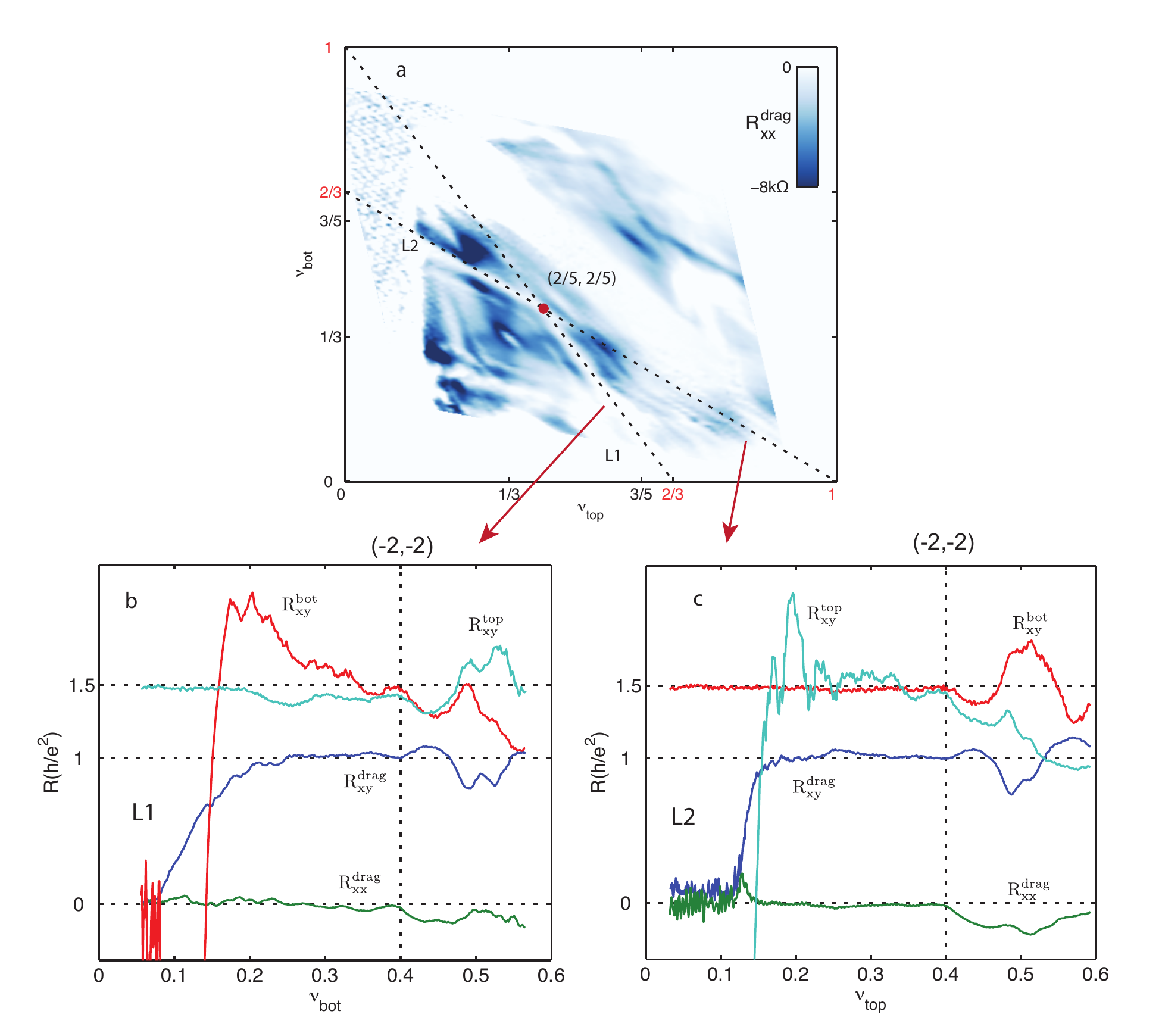}
 \caption{(a) map of longitudinal drag resistance showing where the linecuts are being taken. (b)(c) linecut through $(\nu_{top}, \nu_{bot})=(2/5, 2/5)$ in two different directions. (c) is the same main text Fig. 3f. This is to show that the semi-quantization behavior exists for both linecut direction.}{\label{FS3}}

\end{figure*}

\bibliography{Fractional_drag.bib}